\documentclass[%
 reprint,
superscriptaddress,
 amsmath,amssymb,
 aps,
]{revtex4-2}

\usepackage{graphicx}
\usepackage{dcolumn}
\usepackage{bm}
\usepackage{physics}

\usepackage[dvipsnames]{xcolor}
\usepackage{mathtools}
\usepackage{caption}
\usepackage{subcaption}

\begin{document}

\title{Prediction of topological phases in metastable ferromagnetic MPX$_3$ monolayers}%
\author{Natalya Sheremetyeva}
\affiliation{Thayer School of Engineering, Dartmouth College, Hanover, NH 03755, USA}

\author{Ilyoun Na}
\affiliation{Department of Physics, University of California, Berkeley, CA, 94720, USA}

\author{Anay Saraf}
\affiliation{Thayer School of Engineering, Dartmouth College, Hanover, NH 03755, USA}

\author{Sin\'{e}ad M. Griffin}
\affiliation{%
 Molecular Foundry, Lawrence Berkeley National Laboratory, Berkeley, CA 94720, USA
}%
\author{Geoffroy Hautier}
\affiliation{Thayer School of Engineering, Dartmouth College, Hanover, NH 03755, USA}

\date{\today}

\begin{abstract}
Density functional theory calculations are carried out to study the electronic and topological properties of \textit{M}P\textit{X$_3$} (\textit{M} = Mn, Fe, Co, Ni, and \textit{X} = S, Se) monolayers in the ferromagnetic (FM) metastable magnetic state. We find that FM MnPSe$_3$ monolayers host topological semimetal signatures that are gapped out when spin-orbit coupling (SOC) is included. These findings are supported by explicit calculations of the Berry curvature and the Chern number. The choice of the Hubbard-$U$ parameter to describe the $d$-electrons is thoroughly discussed, as well as the influence of using a hybrid-functional approach. The presence of band inversions and the associated topological features are found to be formalism-dependent. Nevertheless, routes to achieve the topological phase via the application of external biaxial strain are demonstrated. Within the hybrid-functional picture, topological band structures are recovered under a pressure of 15\% (17 GPa). The present work provides a potential avenue for uncovering new topological phases in metastable ferromagnetic phases.
\end{abstract}

\maketitle
\section{\label{sec:intro}Introduction}

Topological insulators and semimetals can exhibit symmetry-protected dissipationless conducting channels on their surface or edges. Such topological features make these quantum materials potential candidates for applications ranging from novel low-energy electronic devices to hosts for Majorana modes for topological quantum computing~\cite{Freedman_et_al:2003}. Finding non-trivial topology in 2D materials is particularly appealing for their potential as low-dimensional device components and their large range of tunability~\cite{Kou_et_al:2017}. The prototypical example of this is graphene~\cite{Wallace:1947}, with many more systems being identified in theory and experiment, and proposed in high-throughput searches~\cite{Choudhary_et_al:2020, Liu_et_al:2018}.

A subset of 2D topological materials are those that break time-reversal symmetry (TRS) giving rise to the Quantum Anomalous Hall Effect (QAHE). QAHE manifests as a quantized Hall conductivity without the need of an external magnetic field and requires long-range ferromagnetic (FM) order with strong spin-orbit coupling (SOC) \cite{Zhang2015}. In three-dimensional (3D) materials, QAHE was first predicted and realized in Cr- or V-doped (Bi,Sb)$_2$Te$_3$ thin films \cite{Chang2013}. In 2D materials, long-range magnetic ordering is rare since strictly it is precluded in rotational invariant systems with short-range magnetic interactions via the Mermin-Wagner Theorem~\cite{Mermin/Wagner:1966}. Recent discoveries of magnetism in 2D materials circumvent this requirement including the discovery of ferromagnetic CrI$_3$ monolayers\cite{Huang_et_al:2017}, and Fe-intercalated TaS$_2$\cite{Husremovic_et_al:2022}. However, finding both non-trivial topology and long-range ferromagnetic order in a 2D material remains particularly challenging.

Theoretical predictions of 2D ferromagnetic topological materials include CoBr$_2$\cite{Chen_et_al:2017}, Fe\textit{X$_3$} (\textit{X} = halide)\cite{Li:2019}, graphene/Cr$_2$Ge$_2$Te$_6$ heterostructures\cite{Zhang2015}, and many more results from high-throughput searches\cite{Choudhary_et_al:2020, Liu_et_al:2018}. However, few experimental signatures of QAHE in any system (3D or 2D) have been observed, and have been limited to very precise synthesis conditions. Therefore, routes to achieving QAHE phases in experimentally realizable materials and conditions are sought after, particularly those in 2D systems. 

Transition metal phosphorus trichalcogenides (TMPTs) are a family of layered 2D magnetic materials with the general formula \textit{M}P\textit{X}$_3$, where \textit{M} is a 3\textit{d} transition metal and \textit{X} is a chalcogen (S or Se). They have recently re-emerged as a particularly interesting family of materials owing to their broad range of tunability and applications ranging from electrochemistry to strongly correlated physics~\cite{Samal_et_al:2021}. In TMPTs, the transition metals form a honeycomb lattice structure and are typically antiferromagnetically coupled in-plane with either a N\'{e}el or striped AFM ground state~\cite{Mak2019}. The AFM bulk and monolayers have been studied extensively for their wide band gaps ranging from 1.3 to 3.5 eV, which makes these semiconductors useful for a variety of optoelectronic applications in a broad wavelength range~\cite{Du2016,Zhang2016}. However, to the best of our knowledge, the FM metastable state of these materials has not been as widely explored. The AFM ordering in monolayer TMPTs has been shown to mostly remain energetically favorable under strain/pressure~\cite{Chittari2016,Pei2018}. Nevertheless, the FM ordering can in principle be achieved via application of a magnetic field or in some cases with carrier doping~\cite{Li2014,Chittari2016}. Stabilizing the FM order can considerably change the electronic properties for the exchange-driven splitting of the bands. For example, topological semimetal features have been predicted to appear in hypothetical FM-ordered multiferroic hexagonal manganites~\cite{Weber2019}.

In this work, we investigated the electronic and topological properties of TMPT monolayers with FM ordering using first-principles calculations based on density functional theory (DFT). The calculated DFT electronic band structures exhibit band crossings near the Fermi level which are gapped out from the introduction of spin-orbit coupling, resulting in an insulating state. The resulting electronic structures are confirmed to have non-trivial topological character via analysis using symmetry indicators and direct calculation of the Berry curvature. We additionally perform a comprehensive stress test of our predictions with  DFT$+U$ and HSE exchange-correlation functionals. We find that though larger $U$ values and HSE open up a considerable gap that results in a topologically trivial state, we can re-engineer the non-trivial phase through moderate amounts of pressure. This study provides a potential new avenue to uncover hidden topological phases in FM metastable states of nominally ground-state AFM systems.

\section{\label{sec:methods}Computational details}
First-principles calculations based on Density-Functional Theory (DFT) were carried out using the Vienna ab initio simulation package (VASP)~\cite{Kohn1965, Kresse1996,Kresse1999}. VASP uses a plane-wave basis set and the projector augmented-wave (PAW) method to describe electron-ion interactions~\cite{Blochl1994}. The Perdew-Burke-Ernzerhof (PBE) generalized gradient approximation (GGA) was adopted for the exchange-correlation potential~\cite{Perdew1996} in the first round of screening calculations. MPX$_3$ monolayers were modeled within a periodic slab geometry. A vacuum region of 24~\AA ~in the out-of-plane direction was used for all slabs to avoid spurious interactions with periodic images. A plane-wave basis energy-cutoff of 520~eV and dense Monkhorst–Pack~\cite{Monkhorst1976} $k$-point grid samplings of the Brillouin Zone (BZ), 13$\times$13$\times$1 (for selenides) and 15$\times$15$\times$1 (for sulfides), were used. Cell parameters and atom positions were optimized until the residual forces were below 2~meV/$\text{\AA}$ while keeping the slab volume fixed with the resulting lattice parameters given in the supporting information (SI) \cite{}[link to SI]. The total energy convergence criterion for the electronic self-consistent loop was set to $10^{-7}\,$eV. After the initial screening, promising candidate materials were also analyzed using the GGA$+U$ method in the Dudarev approach \cite{Dudarev1998} with different $U$ values and with the screened hybrid functional HSE06 \cite{Krukau2006}. For the latter, HSE06 was used only for calculation of the electronic properties, with GGA structures. The topology of the calculated electronic band structures was analyzed using symmetry indicators~\cite{Bradlyn2017,Kruthoff2017,Po2017,Khalaf2018,Song2018} as implemented on the Bilbao Crystallographic Server (BCS) \cite{Aroyo2011,Xu2020,Elcoro2021} and via the explicit calculation of the Berry curvature and the Chern number. For the latter, maximally localized Wannier functions~(MLWFs) were constructed using the Wannier90 code~\cite{Vanderbilt2001_MLWF,Vanderbilt2006_Berry_Curvature_Phase,Vanderbilt2012_MLWF}. For the Wannierization procedure and the construction of the tight-binding Hamiltonian, a DFT calculation with dense $k$-point grid samplings of 18$\times$18$\times$1 was used. 
When applying pressure, the in-plane lattice parameters were scaled by an appropriate factor, and ionic positions were allowed to relax while keeping the cell volume and shape constant.
\subsection{\label{sec:Aresults}Structural and magnetic properties}
\begin{figure}[t]
    \centering
    \includegraphics[width=\columnwidth]{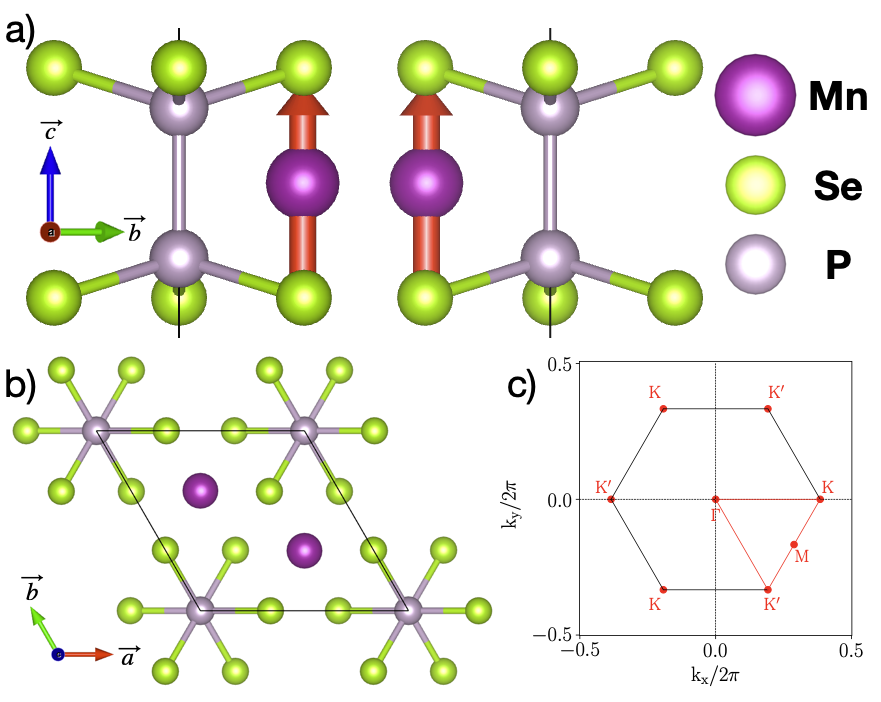}
    \caption{Side view (a) and top view (b) of a TMPT-monolayer primitive cell. MnPSe$_3$ monolayer is used as an example. Magnetic moments on the transition metal atoms are aligned along the out-of-plane direction and are depicted by the (red) arrows. The resulting magnetic state is FM. The VESTA visualization tool~\cite{Momma2011} was used to depict the structures. (c) 2D BZ and the associated high-symmetry $k$-path with and associated symmetry-equivalent points.
    }
    \label{fig:Fig1_Struct_BZ}
\end{figure}
Bulk TMPTs mostly  adopt a monoclinic structure ($C2/m$, No. 12), with the exception of MnPSe$_3$ and FePSe$_3$ that are found in a rhombohedral structure ($R\bar{3}$, No. 148) \cite{Wang2018}, comprising van-der-Waals bonded stacked monolayers. In this work, we focus on one isolated monolayer of TMPT as represented in Fig.~\ref{fig:Fig1_Struct_BZ} by a monolayer MnPSe$_3$. Other TMPT monolayers studied in this work have the same structure, but with the transition metal atoms replaced by either Fe, Co, or Ni which are trigonally coordinated by P, and S or Se. While the magnetic ground state of the TMPT monolayers is known to be AFM\cite{Mak2019}, here we study the metastable ferromagnetic (FM) order where all transition-metal moments are aligned parallel to each other in the out-of-plane direction, as shown in Fig.~\ref{fig:Fig1_Struct_BZ}a. Some of the TMPT monolayers have in-plane (\textit{e.g.} MnPSe$_3$ and NiPS$_3$) and others (\textit{e.g.} FePS$_3$ and MnPS$_3$) have out-of-plane magnetic anisotropy \cite{Mak2019}. Despite being a metastable state, FM order can in principle be achieved via an external magnetic field~\cite{Weber2019}, or through gating\cite{Chittari2016}. 
The calculated structural and magnetic properties of the TMPT monolayers are summarized in table S1 in SI together with experimental values where available. We find that the plain DFT approach (without the Hubbard-$U$ correction) provides good agreement of the calculated and experimental parameters.

\subsection{\label{sec:Bresults}Electronic properties}

\begin{figure*}[t]
\captionsetup{skip=0.25\baselineskip}
\captionsetup[subfigure]{font={small}, skip=-1.5pt, singlelinecheck=off,margin={-4.5cm,0cm}}
\centering
     \begin{subfigure}[b]{0.32\textwidth}                
     \caption{}             
     \label{fig:u0ebs_1a}
         \includegraphics[width=\textwidth]{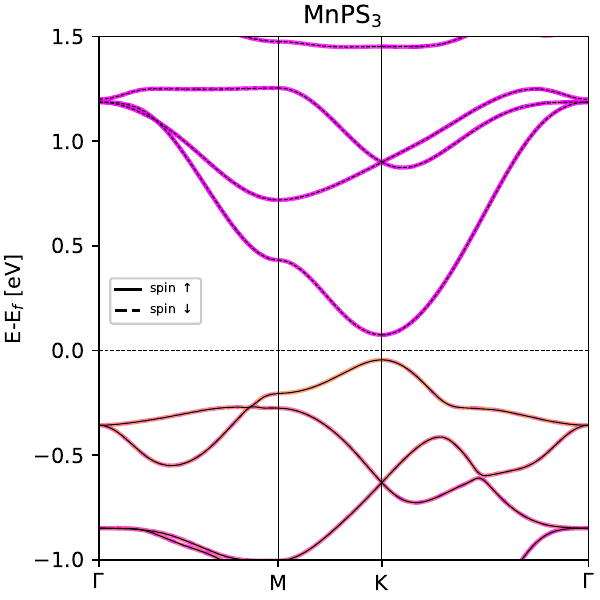}
     \end{subfigure}\hfill%
     \begin{subfigure}[b]{0.32\textwidth}
    \caption{}
             \label{fig:u0ebs_1b}
         \includegraphics[width=\textwidth]{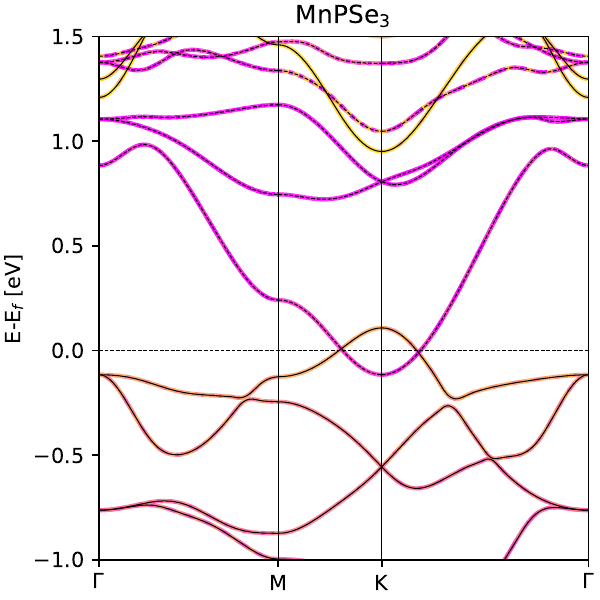}
     \end{subfigure}\hfill%
     \begin{subfigure}[b]{0.32\textwidth}
     \caption{}
                  \label{fig:u0ebs_1c}
         \includegraphics[width=\textwidth]{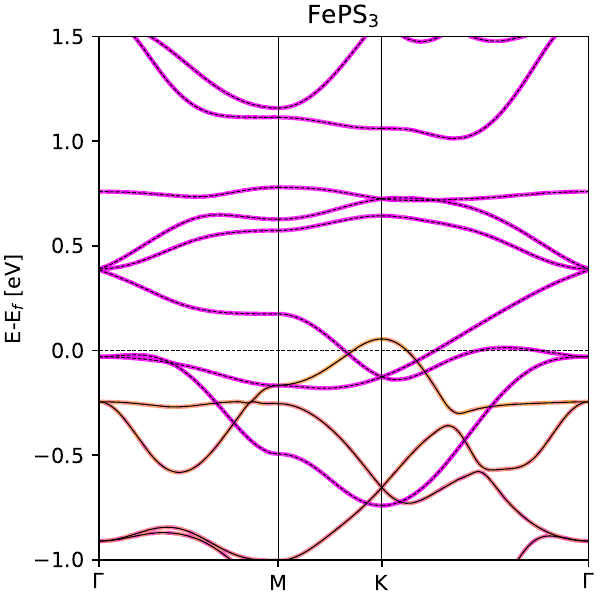}
     \end{subfigure}
     \\\vfill%
     \begin{subfigure}[b]{0.32\textwidth}
     \caption{}
                  \label{fig:u0ebs_1d}
         \includegraphics[width=\textwidth]{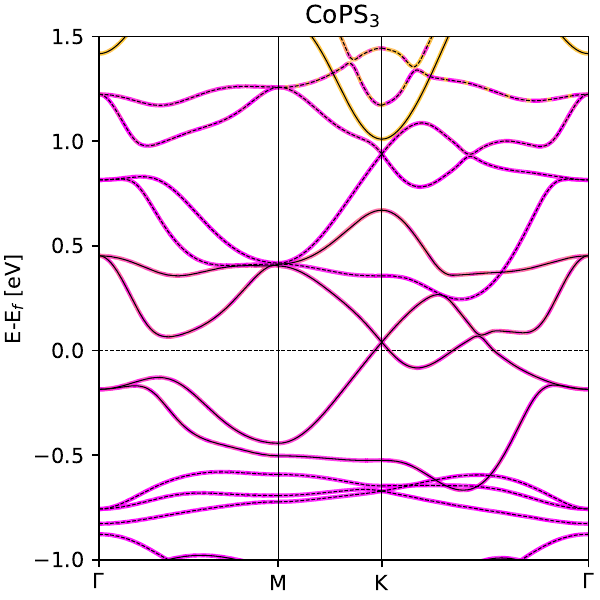}
     \end{subfigure}\hfill%
     \begin{subfigure}[b]{0.32\textwidth}
     \caption{}
                  \label{fig:u0ebs_1e}
         \includegraphics[width=\textwidth]{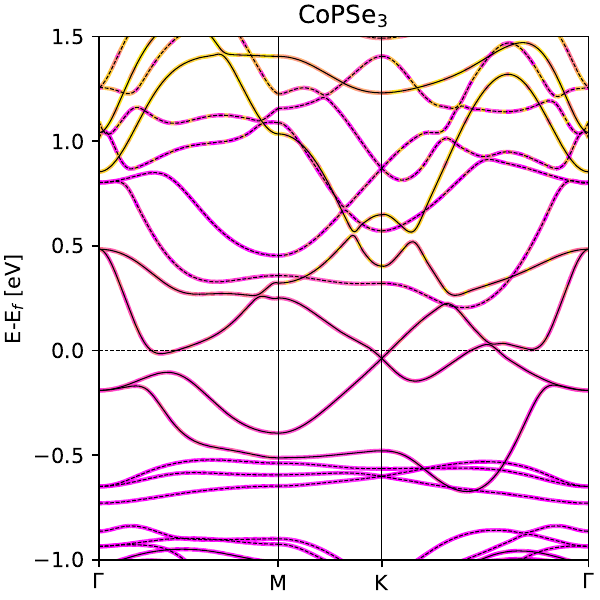}
     \end{subfigure}\hfill%
     \begin{subfigure}[b]{0.32\textwidth}
 \caption{}
              \label{fig:u0ebs_1f}
         \includegraphics[width=\textwidth]{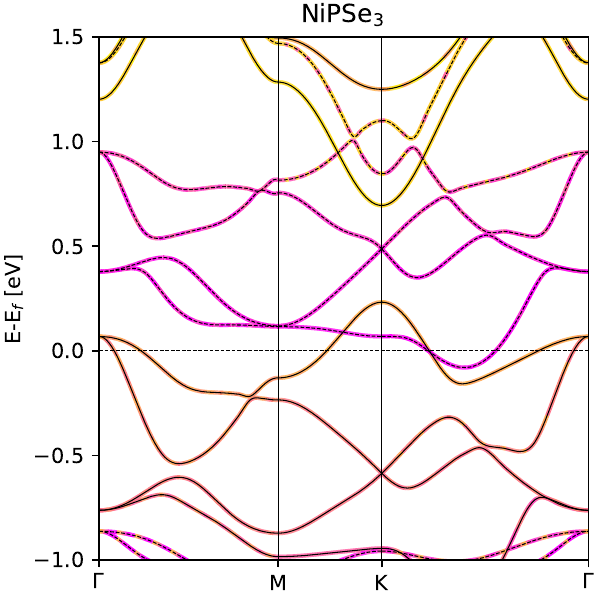}
     \end{subfigure}
     \\
          \begin{subfigure}[b]{0.32\textwidth}
     \end{subfigure}\hfill%
     \begin{subfigure}[b]{0.32\textwidth}
         \includegraphics[width=\textwidth]{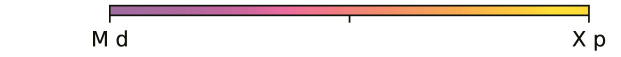}
     \end{subfigure}\hfill%
     \begin{subfigure}[b]{0.32\textwidth}
     \end{subfigure}
     \caption{Orbital-projected GGA ($U$ = 0 eV) electronic band structures without SOC for the ferromagnetic \textit{M}P\textit{X$_3$} compounds (\textit{M} = Mn, Fe ,Co, Ni; \textit{X}= S, Se). Transition metal $d$ states (magenta) and chalcogen $p$ states (yellow) contribute most to the bands near the Fermi level and are used as the orbital-projections range, see the color scale. Spin-up and spin-down bands are marked by solid and dashed black lines, respectively. The horizontal dashed black line at zero energy marks the Fermi level.}
      \label{fig:All_u0ebs}
\end{figure*}
We first calculated the electronic band structures for all combinations \textit{M} and \textit{X} in \textit{M}P\textit{X$_3$} described above with FM order, and their resulting topological classification from symmetry indicators. Fig.~\ref{fig:All_u0ebs} shows the resulting calculated electronic band structures of the topological \textit{M}P\textit{X$_3$} compounds without SOC~\footnote{Fig.~\ref{fig:u0ebs_1a} for MnPS$_3$ does not show topological features but can be made topologically nontrivial by the application of pressure, as will be shown later.} FePSe$_3$ is not included in Fig. \ref{fig:All_u0ebs} as it relaxes to a non-magnetic state \cite{Mak2019} and remains topologically trivial. Similarly, NiPS$_3$ is topologically trivial and thus not included. For NiPS$_3$, the topologically trivial state can be traced back to the prevalent indirect band gap and the associated valence band and conduction band alignment, see SI [link to SI].

With the exception of MnPS$_3$, whose special case will be discussed later, all the band structures in Fig.~\ref{fig:All_u0ebs} are metallic and display possible signatures of non-trivial topology namely (i) degeneracies in the band structures without SOC (that are gapped upon the inclusion of SOC) and (ii) band inversion in these SOC-gapped bands \cite{Yu2010}. Moreover, the crossing bands near the Fermi level have both spin-up and spin-down character, except for CoPS$_3$ and CoPSe$_3$. The latter show Dirac cones at the $K$ point built by two spin-up bands.  The projections on the dominant atomic orbitals are indicated by the color scale. The dominant orbital contributions arise from the transition metal $d$ states (magenta) and the chalcogen $p$ states (yellow) for all compounds. For most systems, the valence-like band participating in the band crossings is a mixture of $d$ and $p$ orbitals, while the conduction-like band is predominantly of $d$ character. For CoPS$_3$ and CoPSe$_3$, the Dirac cones consist of two bands with mixed $d$ and $p$ orbital character, but with larger $d$ contributions.

Among all the band structures in figures \ref{fig:u0ebs_1b}-\ref{fig:u0ebs_1f}, MnPSe$_3$ is the most promising candidate as a high-quality topological insulator once SOC is included as it has the cleanest inverted band structure near the Fermi level. For this reason, we will mostly focus on this system for the remainder of this work.
For all other compounds, we find trivial metallic states with the location of the sought-after band crossings shifted away from the Fermi level. These could potentially be useful if the bands and/or the Fermi level could be reshaped appropriately to achieve a global insulating state (see SI for full calculations with SOC of these less ideal cases).

\subsubsection{\label{sec:Baresults}Topological features in MnPSe$_3$}
\begin{figure}[ht!]
\centering
     \begin{subfigure}[t]{0.43\textwidth}
          \caption{}
      \label{fig:mnpse3_u0ebs_top}
         \centering
        \includegraphics[width=\textwidth]{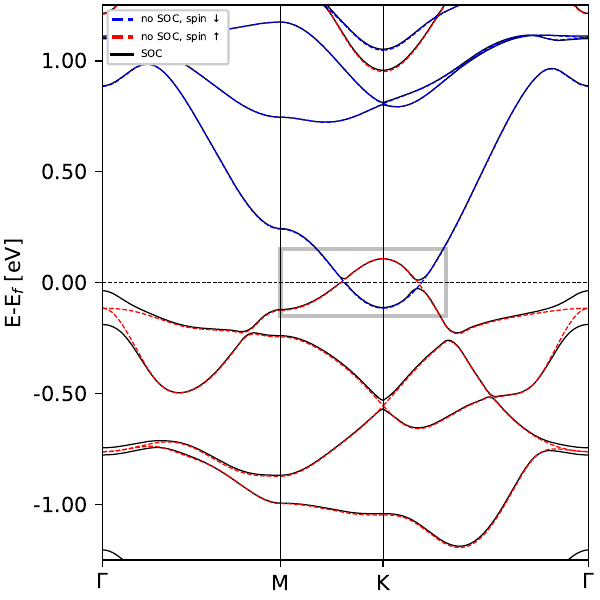}
     \end{subfigure}
     \\
     \begin{subfigure}[t]{0.43\textwidth}
      \caption{}
      \label{fig:ZoomIn_mnpse3_u0ebs}
         \centering
        \includegraphics[width=\textwidth]{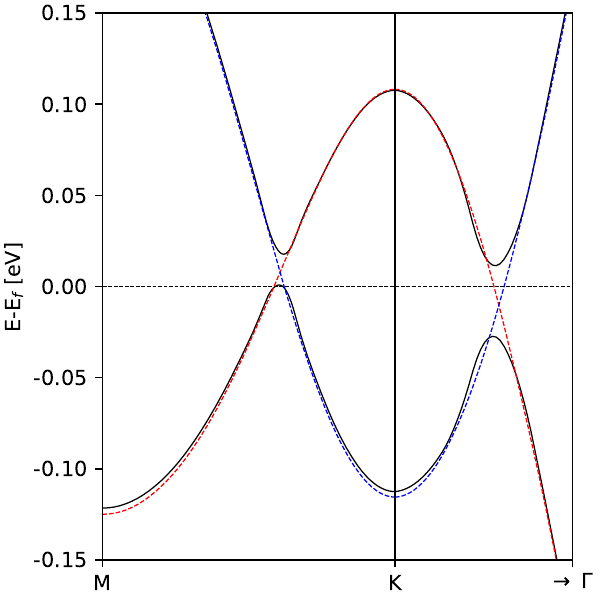}
     \end{subfigure}
          \caption{Electronic band structure of FM MnPSe$_3$ without SOC (red and blue dashed lines for spin-up and spin-down bands, respectively; the bands are the same as in Fig. \ref{fig:u0ebs_1b}) and with SOC (black solid lines). (a) A larger energy window around the Fermi level (marked by the horizontal dashed black line at zero energy). (b) A close-up of the SOC-induced band gap around the $K$ point.
     }
      \label{fig:mnpse3_u0ebs}
\end{figure}
Fig.~\ref{fig:mnpse3_u0ebs} shows the calculated electronic band structure of FM MnPSe$_3$ with and without SOC. As expected, we see that SOC leads to band splittings of previously degenerate bands at different $k$-space points like \textit{e.g} at the $\Gamma$-point slightly below and at the $K$-point about 0.5 eV below the Fermi level $E_f$, in addition to the points in the vicinity of the $K$-point directly at $E_f$. 
Fig.~\ref{fig:ZoomIn_mnpse3_u0ebs}, provides a closer look at the inverted topological bands. We find an asymmetrical (inverted) double-well feature with a smaller SOC-induced direct gap along the $M \rightarrow K$ direction compared to a larger one along the $K \rightarrow \Gamma$ path. This results in an overall indirect SOC-induced bandgap of about 16 meV. 
For a more detailed quantitative evaluation of the SOC-induced band gaps, we calculated the SOC band structure with very dense k-space sampling (560 points per segment), see SI. 
We find the direct gap along $M \rightarrow K$ to be 6 meV, and the direct gap along $K \rightarrow \Gamma$ to be nearly closed ($\sim$0.3 meV) with the CBM lying below $E_f$. Since the exact nature of the gap ($\Delta)$ between these inverted bands is crucial for classifying the material's topology (i.e. semimetal or insulator), we next carefully evaluate the nature of this would-be band gap with respect to symmetry and topological invariants. 

\begin{figure}[t]
    \centering
   \includegraphics[width=\columnwidth]{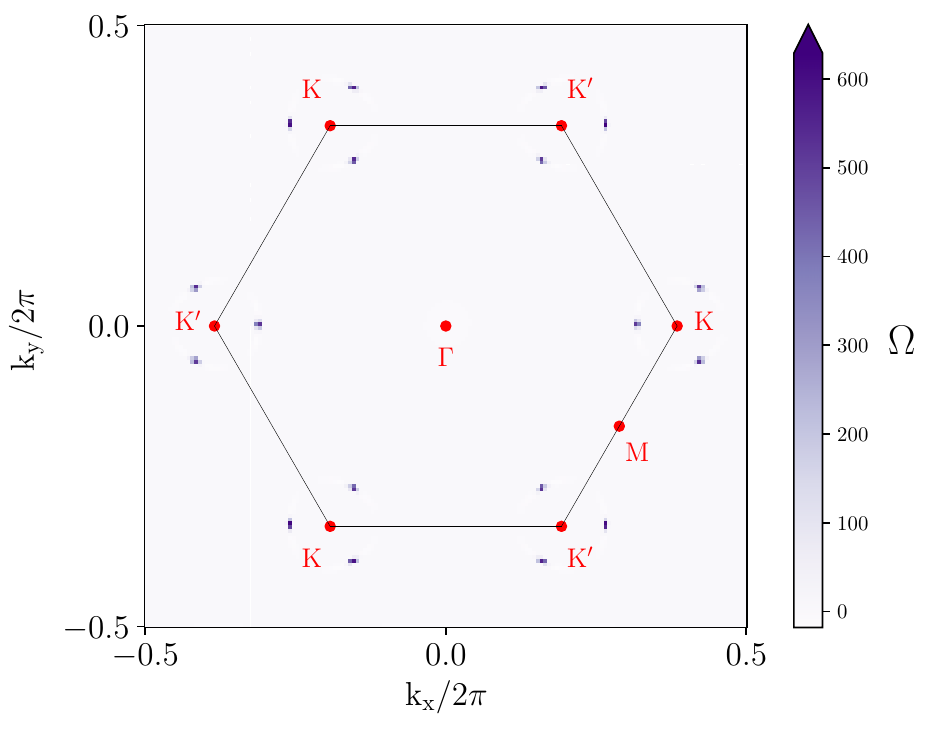}
    \caption{The calculated Berry curvature distribution $\Omega$ (in units of $e^2/h$) of isolated occupying band manifold below the energy gap at $E_F$ in the first BZ and associated symmetry-equivalent points. Large spikes in Berry curvature occur at $k$-points at which very small energy gaps are observed. The integration of the gauge invariant form of the Berry curvature over the BZ gives the Chern number $C=2$. The Chern number of 2 characterizes the non-trivial topology of the isolated band manifold.}
    \label{fig:Fig5_Berry_Curvature}\end{figure}

In the first scenario, where the gap $\Delta$ along  $K \rightarrow \Gamma$  closes, we recover a topological semimetal. This can be additionally classified as either a Dirac (four-fold band degeneracy) or Weyl (two-fold band degeneracy) semimetal in 2D. We next identify which of these two cases we might have for $\Delta = 0$.

The out-of-plane magnetization breaks time-reversal symmetry ($T$) resulting in two inequivalent points $K$ and $K'$ in the BZ. Combining this with the structure's inversion ($P$) and $C_{3z}$ rotation symmetries gives a total of six points near the $K$ and $K'$ points with almost closing gaps, as shown in Fig. S3 in SI. Since these assumed crossings comprise of two bands, these nodal points would be twofold degenerate, resulting in either accidental or symmetry-protected Weyl points~\cite{Kane2015_2DWSM,Gutman2019_2DWSM}. Stabilizing Weyl points in 2D requires additional crystalline symmetries so that the two crossing bands with different symmetry eigenvalues are decoupled. In contrast, in 3D Weyl points are topologically stable without any symmetries as long as either the $P$ or $T$ is broken~\cite{Ashvin2018_WDSM}.

To test for the presence of 2D Weyl points, the Berry phase $\phi$ of the $k$-path loop $l$ enclosing each presumed nodal point is calculated as in~\cite{Zee1984_NonabelianBerry}:
\begin{equation} \label{eq:Berry_Phase_Continuum}
    \phi=\Tr_{\rm occ}{\oint_{l} \vb{A_k}\vdot \vb{dk}}
\end{equation}
where the Berry connection matrix is given by $\vb{A_{m,n}}=i\bra{\Psi_m(\vb{k})}\grad_k \ket{\Psi_n(\vb{k})}$, and the trace is evaluated over occupied bands together with the loop along $k$-space enclosing a nodal point. For a Weyl point, the calculated Berry phase is quantized as $\pi\,(\mathrm{modulo} ~2\pi)$. Here, we calculate that the phase around each presumed nodal point is not quantized, excluding the possibility of a Weyl point. 

Having ruled out the semi-metallic phase ($\Delta=0$), we now investigate the potential topological insulator ($\Delta \neq 0$) phase. The Chern number characterizes the topology of the $N_k$ occupied band manifold  of a system~\cite{Thouless1982_QHC_2D,Niu2010_Berry_RMP}. We calculate the Chern number as the integral of the Berry curvature $\Omega$ over the first BZ. The Berry curvature is used in its gauge invariant form $\Tr_{N_{\vb{k}}}[{\Omega^{z}(\vb{k})}]$, where the trace is taken over the $N_k$ occupied states in the zero-temperature limit. The topology is an intrinsic band property, independent of where the Fermi level sits in a particular system. In our case, the Fermi level passes through the band above the gap in the small region along $K \rightarrow \Gamma$, resulting in  a $k$-dependence of the number of occupied bands $N_k$. Thus, we focus on the band manifolds below the gap along $K \rightarrow \Gamma$. The manifolds can be adiabatically tuned to make the Fermi level cross through the gap without changing the gap's topology as long as the gap is not closing under the symmetry-preserving perturbations.

Fig.~\ref{fig:Fig5_Berry_Curvature} shows the calculated Berry curvature in the first BZ and at the symmetry-equivalent points. At the six unique points of the first BZ, the Berry curvature is strongly peaked. These $k$-points are connected to one another by $P$ and $C_{3z}$ symmetries and are associated with the very small energy band gaps in Fig. S3. The finite $k$-grid expression for the Chern number takes the form as $C=1/(2\pi)\cdot\sum_{\vb{k}}\Tr_{N_{\vb{k}}}[{\Omega^{z}(\vb{k})}]$ and yields $C = 2$. Thus, the band manifolds below the gap along $K \rightarrow \Gamma$ classify FM MnPSe$_3$ as a magnetic Chern insulator.

\subsection{\label{sec:Cresults}Choice of the Hubbard-U parameter}
\begin{figure}[t]
\includegraphics[width=0.43\textwidth]{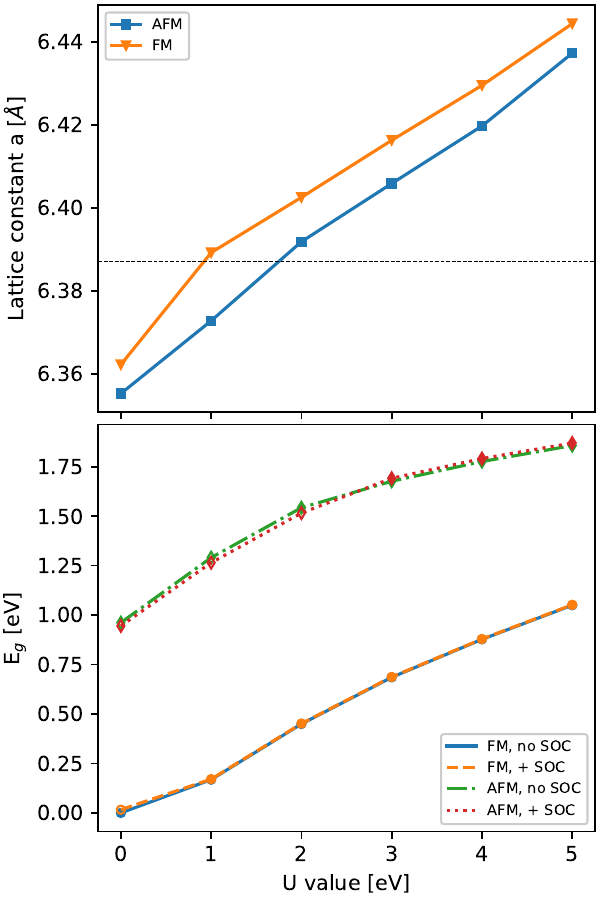}
\caption{\label{fig:Alat_Eg_vs_P_allU} 
Effect of the Hubbard-$U$ parameter on the structural (top) and electronic (bottom) properties of MnPSe$_3$ for the N\'{e}el-AFM ground state and FM metastable state (see legend). Top panel: In-plane lattice constant for the fully relaxed monolayer as a function of $U$ with blue squares and orange triangles corresponding to AFM and FM structures, respectively. The experimental lattice constant of bulk AFM MnPSe$_3$ of 6.387 \AA~ is marked by the horizontal dashed black line for reference~\cite{Wiedenmann1981}. Bottom panel: Similar to the top panel, but for the electronic band gap. Values obtained without and with SOC are shown. Filled markers correspond to zero or direct band gaps, while open markers indicate indirect band gaps.
}
\end{figure}

So far, we have discussed the potential appearance of topological insulator features in FM MnPSe$_3$ with a clean SOC-induced band gap of $\mathcal{O}$(meV). We note, however, that the above discussion was based on the results obtained with GGA. This approximation is known to be inadequate for describing the electronic properties of materials with localized $d$ electrons. In general, the choice of an appropriate $U$ parameter is subject to ongoing research, although, in principle, the $U$ value can be determined self-consistently by using the linear-response approach \cite{Himmetoglu2014}. However, even with the self-consistent $U$-value, the resulting electronic properties can be qualitatively wrong, like for example for few-layer CrGeTe$_3$~\cite{Menichetti_2019}. In our case, the combination of localized $d$ electrons with the more delocalized nature of the semimetallic band structures presents an interesting case for choosing an appropriate $U$ parameter. For MnPSe$_3$ specifically, values as large as 5 eV have been used in the literature~\cite{Pei2018}. In this section, we stress test the topological features in FM MnPSe$_3$ with respect to the $U$-parameter.

In Fig.~\ref{fig:Alat_Eg_vs_P_allU} we show the calculated lattice constant (top panel) and the electronic band gap (bottom panel) as a function of the $U$-parameter for AFM and FM MnPSe$_3$. We find that the calculated lattice constant grows approximately linearly with increasing $U$ for both magnetic states, as has been reported previously for other transition-metal based semimetals~\cite{Griffin/Spaldin:2014}. The horizontal dashed line indicates the experimental lattice constant of bulk MnPSe$_3$ of 6.387 \AA ~\cite{Wiedenmann1981}. For AFM, the best agreement between the computed and the experimental lattice constant is found for $U=2\,$ eV, where the former overestimates the latter by 0.08\%. For all $U$ values, including $U=0$, the calculated AFM lattice constant deviates from the experimental one by less than 1\%. In fact, the largest difference of 0.8\% is found for $U=5\,$ eV, a value previously used in the literature~\cite{Pei2018}. The calculated FM lattice constant is always slightly larger than the AFM one, with the largest difference of 0.3\% for $U=1\,$eV. Given the typical overestimation of the lattice parameters by the GGA functional, our structural results suggest that a zero or small Hubbard-$U$ is most appropriate for comparison with the experiment. However, the structural results do not provide a strong measure for the semi-empirical tuning of the Hubbard$-U$ parameter in the case of MnPSe$_3$.

Turning to the electronic properties, the band gap increases monotonically with $U$ with and without SOC for both magnetic states, as expected for the increased $d$-band localization with $U$. For AFM, the no-SOC gap is indirect for $U=0$ and $U=1\,$ eV and becomes direct for larger $U$ values. The SOC gap is also indirect first and becomes direct starting from $U=2\,$ eV. Interestingly, SOC slightly decreases and increases the band gap for $U\leq2\,$eV and $U>2\,$eV, respectively. Experimentally, a 2.5 eV band gap was reported for bulk MnPSe$_3$ \cite{Du2016}, and a direct gap of 2.32 eV was determined computationally for the monolayer using the HSE06 functional \cite{Zhang2016}. None of the $U$ values result in such a large band gap, although a direct gap can be achieved with at least $U=2\,$eV. The largest gap of 1.86 eV is obtained for $U=5\,$eV, in agreement with a previous GGA$+U$ result \cite{Pei2018}, underestimating the HSE06 gap by 0.5 eV. Thus, the Hubbard$-U$ parameter cannot be selected based on fitting the band gap. Since the Hubbard-$U$ correction only accounts for the underlocalization of the $d$ orbitals and not for excited state corrections, adjusting $U$ to reach the experimental band gap is not necessarily a good benchmark, though it has been successful in previous studies. Finally, the experimental determination of the band gap size and character is strongly technique-dependent, and, to the best of our knowledge, a definitive experimental result for the band gap of AFM MnPSe$_3$ monolayer is still outstanding. 

For FM MnPSe$_3$, a zero band gap and the associated band crossings are found only for $U=0$ without SOC. For all nonzero $U$ values the band gap is finite and direct with and without SOC, resulting in an overall topologically trivial state. The orbital contributions to the valence (VB) and conduction bands (CB) change with increasing $U$, as well. The VB consists of almost exclusively $p$ orbitals and the CB changes to a mixture of $d$ and $p$ orbitals with larger contributions from the latter for $U=5\,$eV, see SI. Despite these changes in orbital contributions, the shapes of VB and CB and the direct alignment of VBM and CBM over each other in $k$-space remain the same for all $U$. Overall, even without a conclusive result on the most appropriate Hubbard$-U$ parameter, the strong $U$-dependence of the previously discussed finding of nontrivial topology requires further investigation.

Returning to the $U$-dependence of the in-plane lattice constant in the top panel of Fig.~\ref{fig:Alat_Eg_vs_P_allU}, it is apparent that increasing $U$ acts as an effective biaxial strain on the structure. At the same time, the band gap grows with $U$, while the shapes of the valence and conduction bands remain mostly unchanged, as already discussed. It is then natural to ask if it is possible to restore the band crossings observed in FM MnPSe$_3$ for $U=0$ by applying external biaxial pressure to the structure when using a nonzero $U$ value. This question can be answered affirmatively as we will discuss in the next section.
\subsection{\label{sec:Dresults}Effect of pressure}
\begin{figure}[t]
\includegraphics[width=0.43\textwidth]{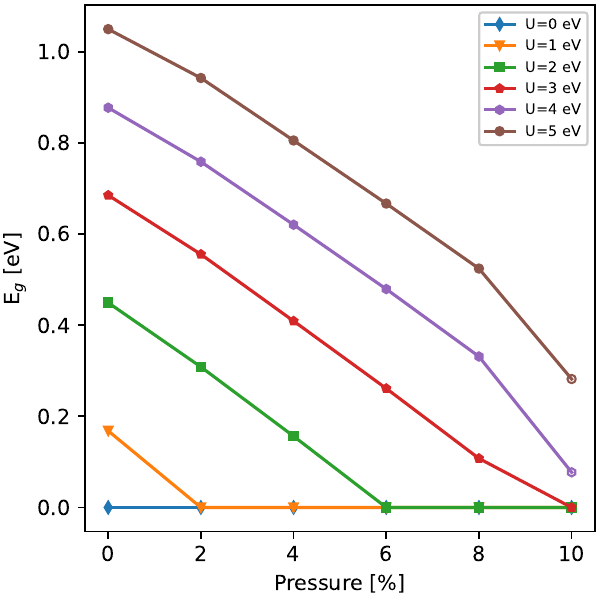}
\caption{\label{fig:Eg_vs_P_allU}
The electronic band gap of the FM MnPSe$_3$ monolayer without SOC as a function of applied external pressure for different values of the Hubbard-$U$ parameter (see legend). The external pressure is given relative to the fully optimized lattice constant for a given $U$. The maximum applied pressure of 10\% corresponds to roughly 4 GPa for all $U$s. Filled markers correspond to zero or direct band gaps, open markers indicate indirect band gaps.
}
\end{figure}
In Fig.~\ref{fig:Eg_vs_P_allU} we plot the evolution of the band gap calculated without SOC for FM MnPSe$_3$ monolayer as a function of the relative biaxial pressure for different $U$ values (see legend). As hypothesized, the band gap decreases with increasing pressure for all $U$ values owing to the increased orbital overlap. The pressure required to close the band gap increases with $U$ owing to the competing effects of the increased localization with increasing $U$, and the greater orbital overlap from increasing pressure. For example, we recover a metal  with only 2\% pressure (0.6 GPa) for $U=1\,$, while for $U=3\,$ eV 10\% pressure (4.2 GPa) is required. For $U=4\,$ and $5$ eV, we never observe a band gap closing, but it appears possible with the downward trend of $E_g\left(U\right)$. Furthermore, the band gap remains direct for all pressures before becoming metallic for $U\leqslant3\,$eV. In contrast, for $U=4\,$ and 5 eV, the band gap becomes indirect at $P=10\%$ with the VBM now at $\Gamma$ instead of at $K$. Nevertheless, the valence band extremum at $K$ remains prominent, making band-inverted crossings in that momentum-space region in principle possible under pressure, see the associated band structures in SI [link to SI].
\begin{figure*}[t]
\captionsetup[subfigure]{justification=centering}
\centering
     \begin{subfigure}[t]{0.44\textwidth}
         \centering
         \caption{\label{fig:mnpse3_u3p0ebs}$U=0\,$eV, $P=$0\% (0 GPa)}
         \includegraphics[width=\textwidth]{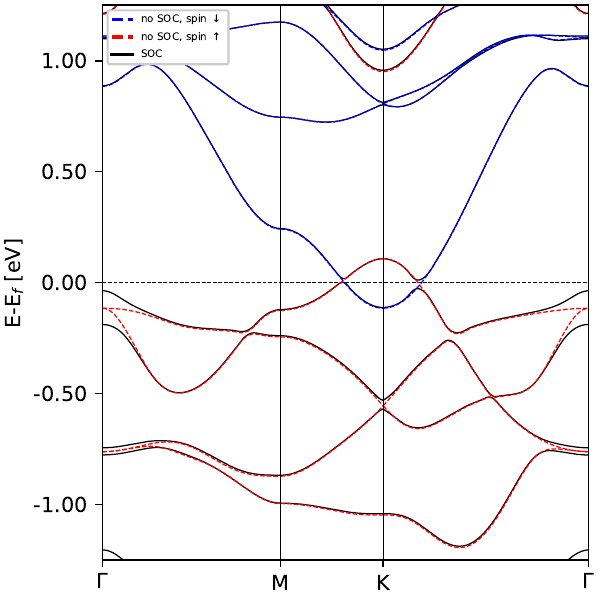}
     \end{subfigure}
     \hfill
          \begin{subfigure}[t]{0.44\textwidth}
         \centering
         \caption{\label{fig:mnpse3_u3p10ebs}$U=3\,$eV, $P=$10\% (4 GPa)}
         \includegraphics[width=\textwidth]{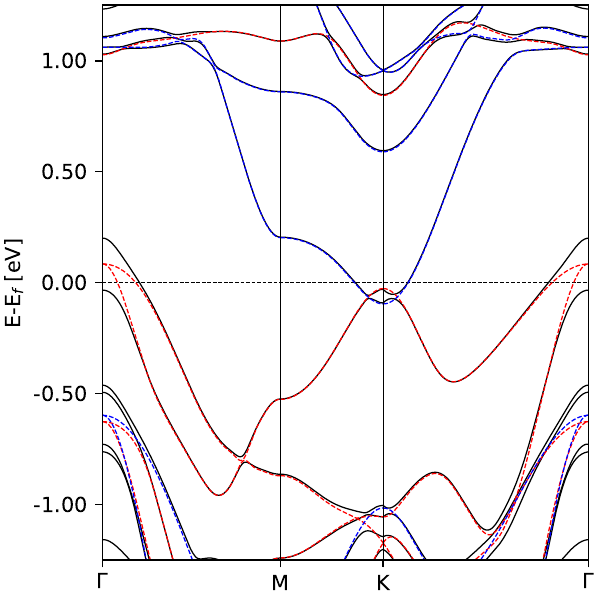}
     \end{subfigure}
          \\
     \begin{subfigure}[t]{0.44\textwidth}
         \centering
         \caption{\label{fig:mnpse3_u5p16ebs}\centering{$U=5\,$eV, $P=$16\% (8 GPa)}}
         \includegraphics[width=\textwidth]{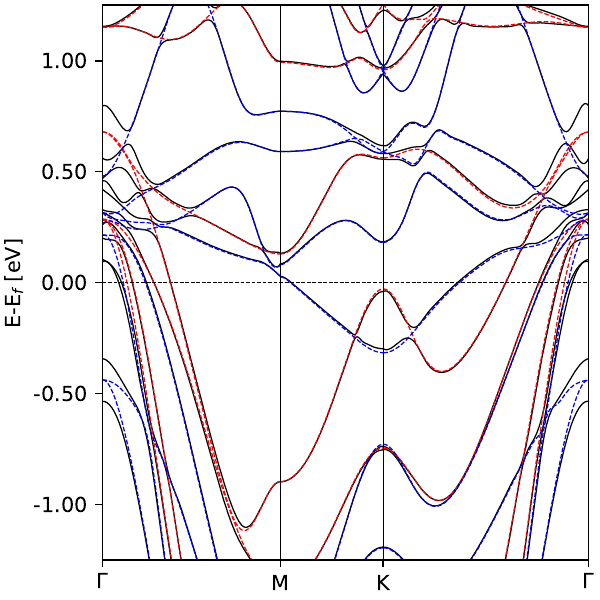}
     \end{subfigure}
     \hfill
     \begin{subfigure}[t]{0.44\textwidth}
         \centering
         \caption{\label{fig:HSE_mnpse3_ebs}HSE06, $P=$15\% (17 GPa)}
         \includegraphics[width=\textwidth]{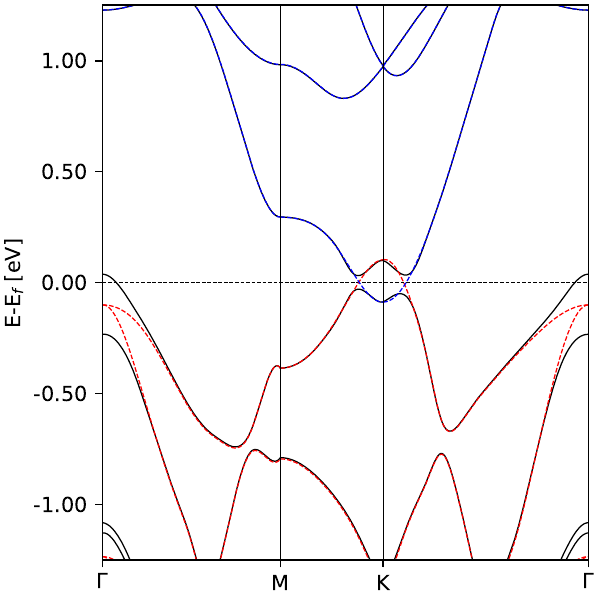}
     \end{subfigure}
          \caption{Similar to figure \ref{fig:mnpse3_u0ebs}, but collecting band structures from different calculation methods (different $U$ values and HSE) and different applied pressures at which a band inversion centered at $K$ is obtained with each method. a) Same as Fig.~\ref{fig:mnpse3_u0ebs} for reference. b) Using $U=3\,$ eV and under 10\% (4 GPa) external pressure. c) Using $U=5\,$ eV and under 6\% (8 GPa) external pressure. d) Using the HSE06 functional and applying 15\% (17 GPa) pressure.
     }
      \label{fig:U3P10_U5P14_ebs}
\end{figure*}
The effect of SOC is included in Fig.~\ref{fig:U3P10_U5P14_ebs} where we plot the calculated band structures for two representative $U$ values at the corresponding pressures necessary for FM MnPSe$_3$ to have a band crossing without including SOC. For $U=3\,$ eV, the required pressure is about 10\% (4 GPa). In fact, the band structure without SOC becomes metallic at $P=9\%$ due to the minimum of the conduction-like band at $K$ being shifted slightly below $E_f$, but the band inversion has not yet occurred at that pressure. At $P=10\%$, the band-inversion crossing appears about 60 meV below the Fermi level. At the same time, the maximum of the valence-like band at $\Gamma$ is about 85 meV above $E_f$. Upon introduction of SOC, the band crossing points around $K$ split with a small gap which is topologically protected from our calculation of the symmetry indicators. The degeneracy of the valence-like band at $\Gamma$ is also lifted by SOC with one of the bands having its maximum now about 34 meV below and the other about 200 meV above the Fermi level. As a result, the entire band structure in Fig. \ref{fig:mnpse3_u3p10ebs} remains metallic even after SOC is included. \\
For $U=5\,$ eV, the pressure required to achieve band inversion is about 16\% (8 GPa). Overall, the band structure in Fig. \ref{fig:mnpse3_u5p16ebs} is qualitatively more complicated than that for $U=3\,$ eV with $P=10\%$, having many bands around the $\Gamma$ point crossing $E_f$. Similarly to $U=3\,$ eV, the band structure becomes metallic already at $P=12\%$, before the band crossings appear, due to the VBM at $\Gamma$ and CBM at $K$ shifting towards $E_f$. At $P=16\%$, the band crossing is located at about 250 meV below the Fermi level and is split by SOC into a small gap located somewhat of the $K$ point along the $M\rightarrow K$ path. The second band splitting along the $K\rightarrow \Gamma$ path is larger with a magnitude of $\sim$250 meV.

\subsubsection{\label{sec:Daresults}Topological features in MnPSe$_3$ under pressure with HSE functional}
Our results thus far show the sensitivity of our calculated electronic structure, and hence topological character, on the choice of exchange-correlation functional. In particular, we find that while the structural parameters do not have a strong dependence on our Hubbard-$U$ parameter, there is a wide variation in the predicted band gaps and the corresponding presence/absence of band inversion and nontrivial topology. We next explore the use of hybrid functional approaches which can be more accurate for describing unoccupied states and band gaps but come with an additional computational cost. 

Here, we have used the HSE06 functional for both AFM and FM orderings of the MnPSe$_3$ monolayer. The HSE calculation was performed using the GGA-relaxed structure since we saw little variation in structural parameters with functional choice (\textit{i.e.} no structural optimization was performed with HSE). First, the previously reported direct band gap of 2.38 eV at the $K$ point of the first BZ was reproduced for the AFM monolayer \cite{Zhang2016}. As mentioned before, this value is 0.5 eV larger than the one calculated with the largest $U=5\,$eV used in the literature \cite{Pei2018}. Second, we obtained a direct gap of 1.30 eV at the $K$ point for the FM monolayer. This value is similarly 0.25 eV above the one obtained with $U=5\,$eV. The orbital projections of the VB and CB are practically the same for HSE and $U=5\,$eV that were discussed in the previous section, see SI. If we assume that the hybrid functional gives the more accurate calculated electronic structure, we next investigate if it is possible to arrive at a topologically nontrivial state via pressure in the HSE picture.

Fig.~\ref{fig:HSE_mnpse3_ebs} shows the electronic band structure of FM MnPSe$_3$ monolayer under 15\% (17 GPa) biaxial pressure. Without SOC, the band structure strongly resembles the one in Fig.~\ref{fig:mnpse3_u0ebs_top} for GGA without pressure with the characteristic band crossings directly at the Fermi level and centered at $K$. Moreover, the orbital projections of the band structures in Fig.~\ref{fig:mnpse3_u0ebs_top} and Fig.~\ref{fig:HSE_mnpse3_ebs} are very similar with the conduction-like band consisting almost exclusively of $d$ orbitals. In Fig.~\ref{fig:HSE_mnpse3_ebs}, the valence-like band at $\Gamma$ is about 100 meV below $E_f$ leaving a clean Fermi level without SOC. Including SOC has a more pronounced effect on the HSE band structure compared to GGA: a gap of about 60 meV opens up at the band crossing points around $K$ without SOC. At the same time, SOC has also a larger effect on the band splitting at $\Gamma$ with one band now being elevated 38 meV above $E_f$, obstructing the clean topological insulating Fermi surface. Nevertheless, given the small energy difference between the CBM near $K$ and the VBM at $\Gamma$, FM MnPSe$_3$  monolayer under 15\% strain might display topological insulating behavior in practice and with an improved topological band gap than was originally found in GGA or GGA+$U$ picture.

\subsubsection{\label{sec:Dbresults}Topological features in MnPS$_3$}
\begin{figure}[t]
\centering
      \label{fig:mnps3_U3P14ebs_top}
         \centering
         \includegraphics[width=0.43\textwidth]{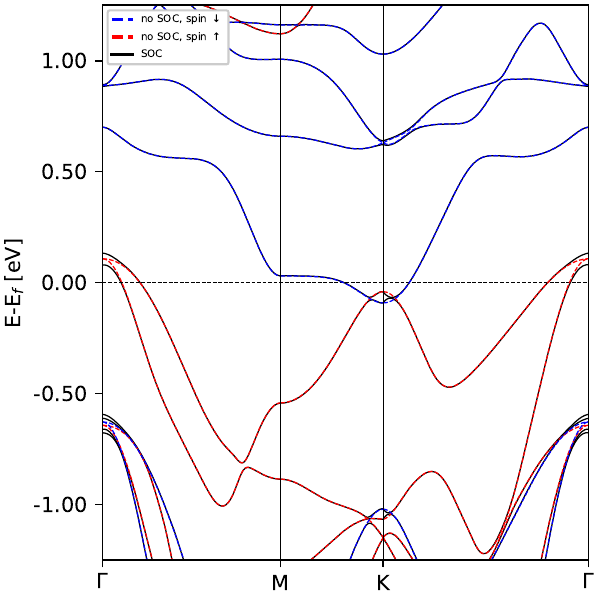}
          \caption{Similar to figure \ref{fig:U3P10_U5P14_ebs}, but for FM MnPS$_3$ using $U=3\,$ eV and applying 14\% biaxial pressure (about 8 GPa).
     }
      \label{fig:mnps3_U3P14ebs}
\end{figure}
So far, we have examined the effect of pressure on the electronic band structure of MnPSe$_3$. It was shown that it is possible to achieve a band inversion for any Hubbard-$U$ parameter via application of an appropriate pressure. We find that a helpful starting point for finding a pressure-induced band inversion is a band structure with directly aligned valence(-like) and conduction(-like) bands. Earlier, we mentioned that MnPS$_3$ shows a direct gap, see Fig.~\ref{fig:u0ebs_1a}. This makes MnPS$_3$ a good candidate for displaying topological features under pressure. Fig.~\ref{fig:mnps3_U3P14ebs} shows the calculated band structures of FM MnPS$_3$ under 14\% (about 8 GPa) biaxial pressure calculated using $U=3\,$ eV. Qualitatively, the band structures are similar to those of MnPSe$_3$ in Fig.~\ref{fig:mnpse3_u3p10ebs}. Indeed, band crossings centered at the $K$ point occur. SOC induces a splitting of 9 meV at the crossing points, which is smaller than in MnPSe$_3$ under similar conditions, as expected for the lighter S atoms.

\section{\label{sec:conc}Conclusion}
We have studied the electronic properties of hypothetical FM MPX$_3$ monolayers to identify possible topologically nontrivial features. We found the most promising candidates to be MnPSe$_3$, and MnPS$_3$ which were then further analyzed using different levels of DFT complexity and characterization of the topological invariants. 
We predict that FM MnPSe$_3$ with out-of-plane magnetization is a magnetic Chern insulator based on  calculations of the Chern number using a Wannier-based tight-binding Hamiltonian.
 The topological band structure in FM MnPSe$_3$ was found to be highly sensitive to the level of the theoretical description. We systematically studied the effect of the Hubbard-$U$ parameter on the electronic properties and also compared to results obtained with HSE06. We found that topological band crossings can be achieved with any formalism under application of appropriate biaxial pressure. Assuming that the topological band structure can indeed be realized in the FM MnPSe$_3$ monolayer, we now discuss potential difficulties of its measurement.

First, as mentioned earlier, the magnetic ground state of the MnPSe$_3$ monolayer is N\'{e}el-AFM \cite{Mak2019}. The FM-AFM energetics are also relatively stable under application of small strains/pressures (within 5\%) with the trend of AFM order being increasingly stabilized by pressure \cite{Chittari2016,Pei2018}. Thus, to achieve the FM ordering in practice, an external magnetic field is required. 

Second, the AFM MnPSe$_3$ monolayer is one of a few magnetic layers with in-plane easy-axis \cite{Mak2019}. Here, we also confirmed the in-plane  easy axis for FM order. The easy-axis' orientation is intricately connected to the nontrivial topology, as it governs the magnetic symmetry of the structure. Specifically, in this work an out-of-plane alignment of the magnetic moments (magnetic space group 162.77) was used. Using the in-plane orientation reduces the magnetic symmetry (magnetic space group 12.62). This symmetry reduction results in trivial topology from symmetry indicators. Thus, an applied external field would have to be large enough not only to align all the magnetic moments parallel with each other to achieve an FM$\rightarrow$AFM transition, but also to flip the orientation of the magnetic moments to the out-of-plane direction. 

Finally, we note that such 2D SOC-induced small band gap materials have practical applications as novel dark matter detectors~\cite{Inzani2021}. In fact, the inherent anisotropy of the electronic structure in such 2D materials could provide improved prospects for directional detection based on the scattering of electrons across small band gaps~\cite{Hochberg_et_al:2018, Geilhufe_et_al:2020}.

All in all, provided practical challenges can be overcome, exploring FM orderings of known AFM monolayers might provide a new pathway towards exploiting topologically nontrivial features in these materials. More generally, applying pressure/strain to FM materials with dispersed (as opposed to flat) and aligned valence and conduction bands might reveal topologically nontrivial materials that have been thus far only known as topologically trivial.

\begin{acknowledgments}
This work was funded by the US Department of Energy, Office of Science, Office of Basic Energy Sciences, Materials Sciences and Engineering Division under Contract No. DE-AC02-05-CH11231 (Materials Project program KC23MP). Computational resources were provided by the National Energy Research Scientific Computing Center and the Molecular Foundry, DOE Office of Science User Facilities supported by the Office of Science, U.S. Department of Energy under Contract No. DEAC02-05CH11231. This work has also been made possible by the financial support of the Neukom Institute for Computational Science at Dartmouth College. 
\end{acknowledgments}

%

\end{document}